\documentclass[preprint,aps]{revtex4}
\usepackage{graphicx}
\usepackage{dcolumn}
\usepackage{bm}
\usepackage{ae}
\usepackage{aecompl}

\begin{document}

\preprint{}

\title{Pressure dependence of the magnetoresistance oscillations spectrum of
$\beta$''-(BEDT-TTF)$_4$(NH$_4$)[Fe(C$_2$O$_4$)$_3$]$\cdot$DMF}

\author{Alain Audouard$^{1\dagger}$,  Vladimir N. Laukhin$^{2,3}$, J\'{e}r\^{o}me B\'{e}ard$^{1}$, David Vignolles$^{1}$, Marc Nardone$^{1}$\\
Enric Canadell$^{3}$, Tatyana G. Prokhorova$^{4}$ and Eduard B.
Yagubskii$^{4}$ }

\affiliation {$^1$ Laboratoire National des Champs Magn\'{e}tiques
Puls\'{e}s (UMR CNRS-UPS-INSA 5147), 143 avenue de Rangueil, 31400
Toulouse, France. \\$^2$ Instituci\'{o} Catalana de Recerca i
Estudis Avan\c{c}ats (ICREA), 08010 Barcelona, Spain.\\$^3$
Institut de Ci\`{e}ncia de Materials de Barcelona (ICMAB - CSIC),
Campus UAB, 08193 Bellaterra, Catalunya,
Spain\\
$^4$ Institute of Problems of Chemical Physics, Russian Academy of
Sciences, 142432 Chernogolovka, MD, Russia.}

\date{\today}

\begin{abstract} The pressure dependence of the interlayer magnetoresistance of the quasi-two
dimensional organic metal
$\beta$''-(BEDT-TTF)$_4$(NH$_4$)[Fe(C$_2$O$_4$)$_3$]$\cdot$DMF has
been investigated up to 1 GPa in pulsed magnetic fields up to 55
T. The Shubnikov-de Haas oscillations spectra can be interpreted
on the basis of three compensated orbits in all the pressure range
studied, suggesting that the Fermi surface topology remains
qualitatively the same as the applied pressure varies. In
addition, all the observed frequencies, normalized to their value
at ambient pressure, exhibit the same sizeable pressure
dependence. Despite this behavior, which is at variance with that
of numerous charge transfer salts based on the BEDT-TTF molecule,
non-monotonous pressure-induced variations of parameters such as
the scattering rate linked to the various detected orbits are
observed.

\end{abstract}


\maketitle

\begin{figure}                                                      
\centering
\resizebox{0.8\columnwidth}{!}{\includegraphics*{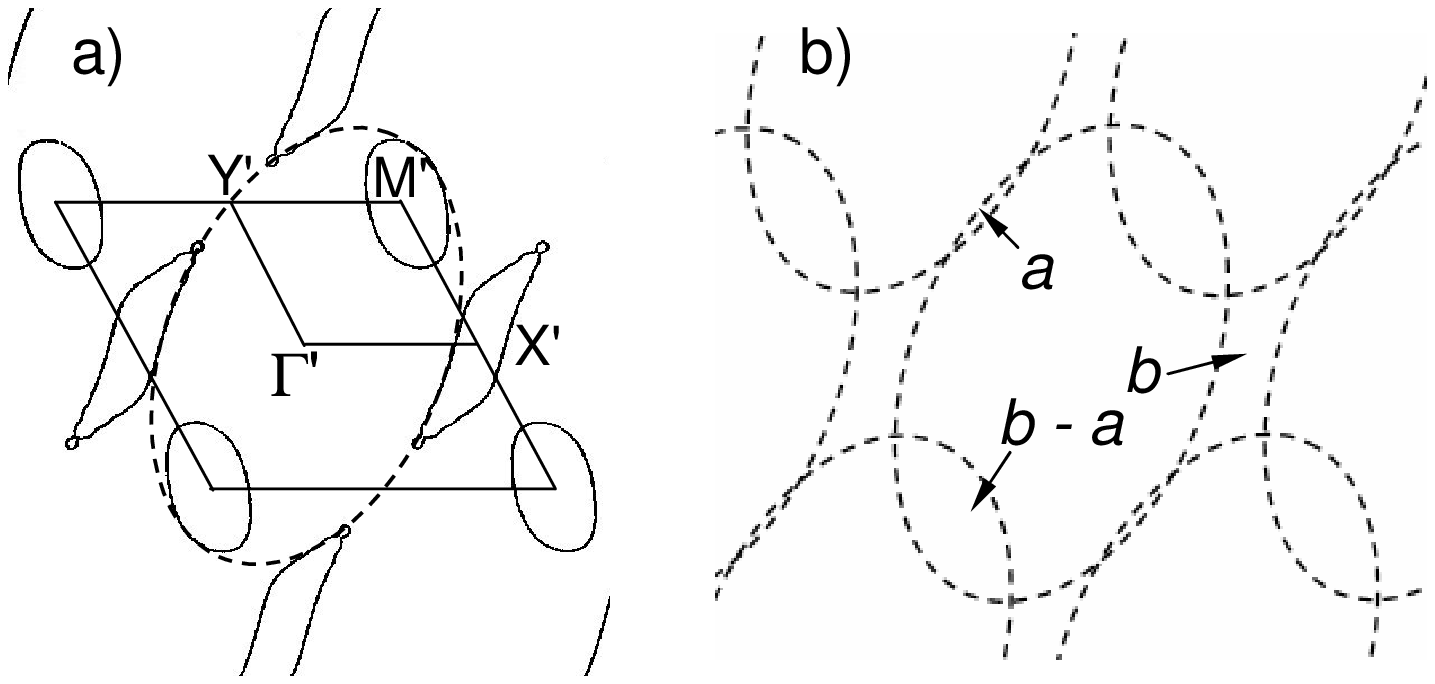}}
\caption{\label{fig_sf} (a) Fermi surface of
$\beta$''-(BEDT-TTF)$_4$(NH$_4$)[Fe(C$_2$O$_4$)$_3$]$\cdot$DMF at
ambient pressure according to band structure calculations
\cite{Pr03}. (b) Representation of intersecting elliptic hole
orbits ($\bigodot$ orbits in dashed lines) leading to three
compensated electron ($b$) and hole ($a$ and $b - a$) orbits. The
area of the $\bigodot$ orbits is equal to that of the first
Brillouin zone (see text).}
\end{figure}

The family of isostructural monoclinic charge-transfer salts
$\beta$''-(BEDT-TTF)$_4$(A)[M(C$_2$O$_4$)$_3$]$\cdot$Solv have
received much attention since it yielded, more than ten years ago,
the first organic superconductor at ambient pressure with magnetic
ions \cite{Gr95}. In the above formula, BEDT-TTF stands for the
bis(ethylenedithio)tetrathiafulvalene molecule, A is a monovalent
cation (A = H$_3$O$^+$, K$^+$, NH$_4$$^+$, etc.), M is a trivalent
cation (M = Cr$^{3+}$, Fe$^{3+}$, Ga$^{3+}$, etc.) and Solv is a
solvent molecule such as benzonitrile, dimethylformamide,
nitrobenzene and pyridine, labelled hereafter BN, DMF, NB and P,
respectively. In the following, these compounds are referred to as
A-M$\cdot$Solv. Even though all these compounds exhibit a metallic
conductivity around room temperature, their electronic properties
strongly depend on subtle details of their molecular arrangement.
In this respect, the disorder which is strongly sensitive to the
nature of the solvent molecules, likely plays a significant role
\cite{Tu99, Ak02}. As an example, whereas H$_3$O-Fe$\cdot$BN is
superconducting with T$_c$ = 8.5 K \cite{Gr95}, H$_3$O-Fe$\cdot$P
exhibits a metal-insulator transition at 116 K \cite{Tu99}.

According to band structure calculations, the Fermi surface (FS)
of H$_3$O-Fe$\cdot$BN \cite{Ku95} and NH$_4$-Fe$\cdot$DMF
\cite{Pr03} [see Fig. \ref{fig_sf}(a)] originates from one
elliptic orbit, of which the cross section area is equal to that
of the first Brillouin zone (FBZ). These orbits overlap  in the
$\Gamma$'M' direction and come into contact at the Y' point which
should yield one electron and one hole compensated orbits with
cross section area of a few percent of the FBZ area, located
around the points X' and M' of the FBZ, respectively.
Nevertheless, the Shubnikov-de Haas (SdH) oscillation spectra
recorded on NH$_4$-Fe$\cdot$DMF \cite{Au04} can rather be
interpreted assuming, as suggested in Ref. \onlinecite{Pr03}, that
overlapping also occurs in the $\Gamma$'Y' direction leading to
two hole and one electron compensated orbits labelled $a$, $b - a$
and $b$, respectively in Fig. \ref{fig_sf}(b). However, this
picture cannot hold for several compounds of the considered
family. Indeed, only two frequencies were observed for
H$_3$O-M$\cdot$NB \cite{Ba05} whereas, four frequencies were
reported for the H$_3$O-M$\cdot$P (M = Cr, Ga, Fe) salts
\cite{Co04}. In these two latter cases, a density wave state,
responsible for the observed strongly non-monotonous temperature
dependence of the resistance, has been invoked in order to account
for this discrepancy. The FS of NH$_4$-Cr$\cdot$DMF is probably
even more complex since the SdH oscillations spectra can be
accounted for by up to six orbits \cite{Vi06} even though a
metallic conductivity is observed down to low temperature.
Nevertheless, applied hydrostatic pressure has a drastic effect on
the FS of this latter compound since the number of Fourier
components involved in the SdH oscillations spectra progressively
decreases down to three as the pressure increases up to 1 GPa. In
addition, these three frequencies are linked by a linear relation
of the form F$_b$ = F$_{b-a}$ + F$_a$. In other words, the FS of
NH$_4$-Cr$\cdot$DMF under pressure would be qualitatively the same
as that of NH$_4$-Fe$\cdot$DMF at ambient pressure.\\

The aim of this paper is to investigate the pressure dependence of
the interlayer magnetoresistance of the NH$_4$-Fe$\cdot$DMF salt.
A behavior strongly different from that of the related compound
NH$_4$-Cr$\cdot$DMF and numerous salts based on the BEDT-TTF
molecule, is observed.\\

The studied crystal was an elongated hexagonal platelet with
approximate dimensions (0.4 $\times$ 0.2 $\times$ 0.1)~mm$^3$, the
largest faces being parallel to the conducting \emph{ab}-plane.
Magnetoresistance experiments were performed in pulsed magnetic
field up to 55 T  with a pulse decay duration of 0.32 s, in the
temperature range from 1.6 K to 4.2 K. Quasi-hydrostatic pressure
was applied in an anvil cell designed for isothermal measurements
in pulsed magnetic fields \cite{Na01}, up to (0.98 $\pm$ 0.05) GPa
at low temperature. The organic liquid GKZh-94 was used as
pressure-transmitting medium. The quoted pressure values at low
temperature are corrected taking into account the decrease of
pressure on cooling. Experimental details for interlayer
resistance measurements and Fourier analysis can be found in Refs.
\cite{Au04,Vi06}.

\begin{figure} [h]                                                     
\centering \resizebox{0.8\columnwidth}{!}
{\includegraphics*{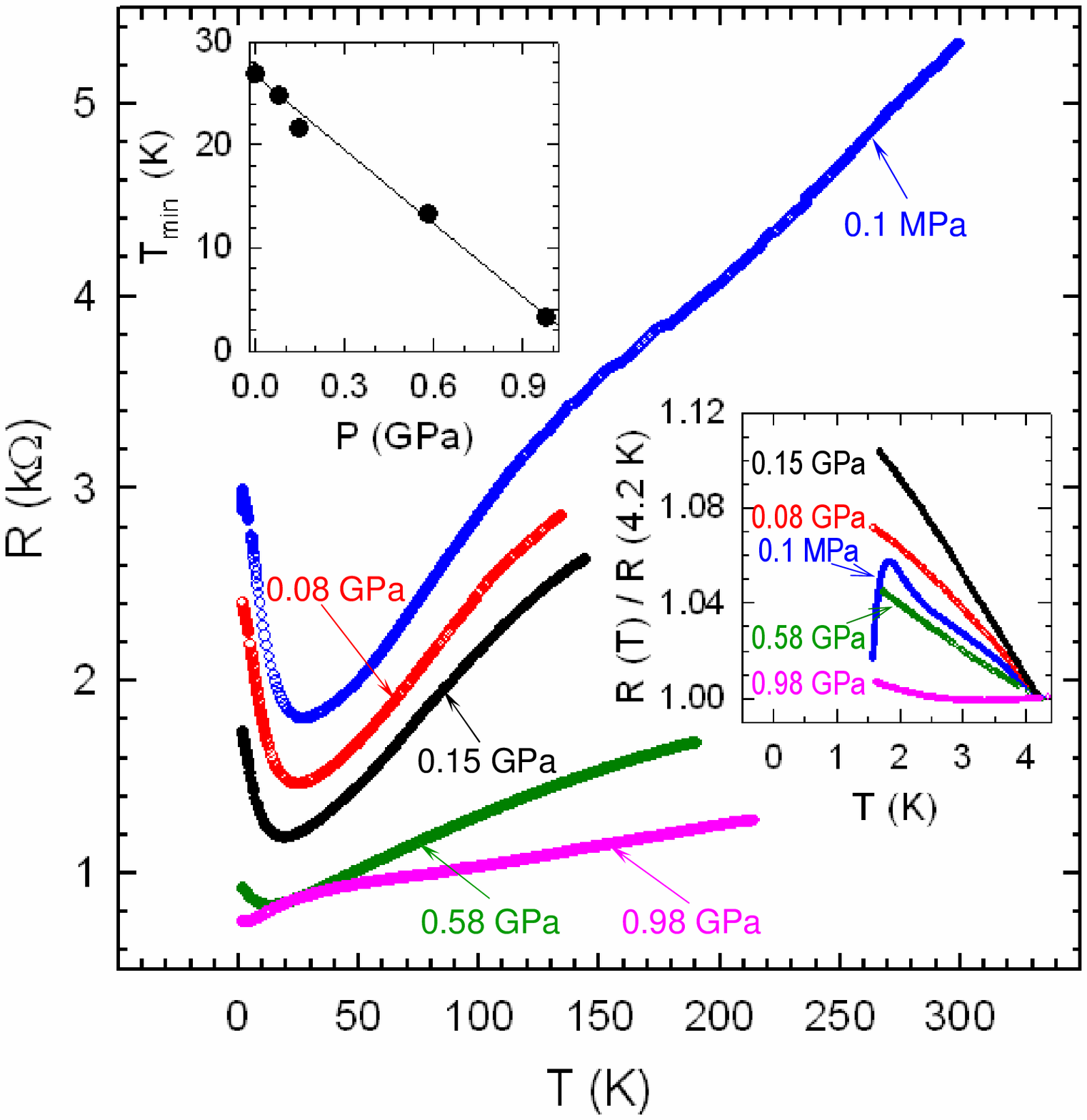}} \caption{\label{fig_rt} (Color online)
Temperature dependence of the zero-field interlayer resistance for
the various pressures applied. The lower inset displays the low
temperature part of the data. The pressure dependence of the
temperature at which the resistance goes to a minimum (T$_{min}$)
is displayed in the upper inset. The pressures applied at low
temperature are indicated in the figure.}
\end{figure}

\begin{figure} [h]                                                     
\centering \resizebox{1\columnwidth}{!}
{\includegraphics*{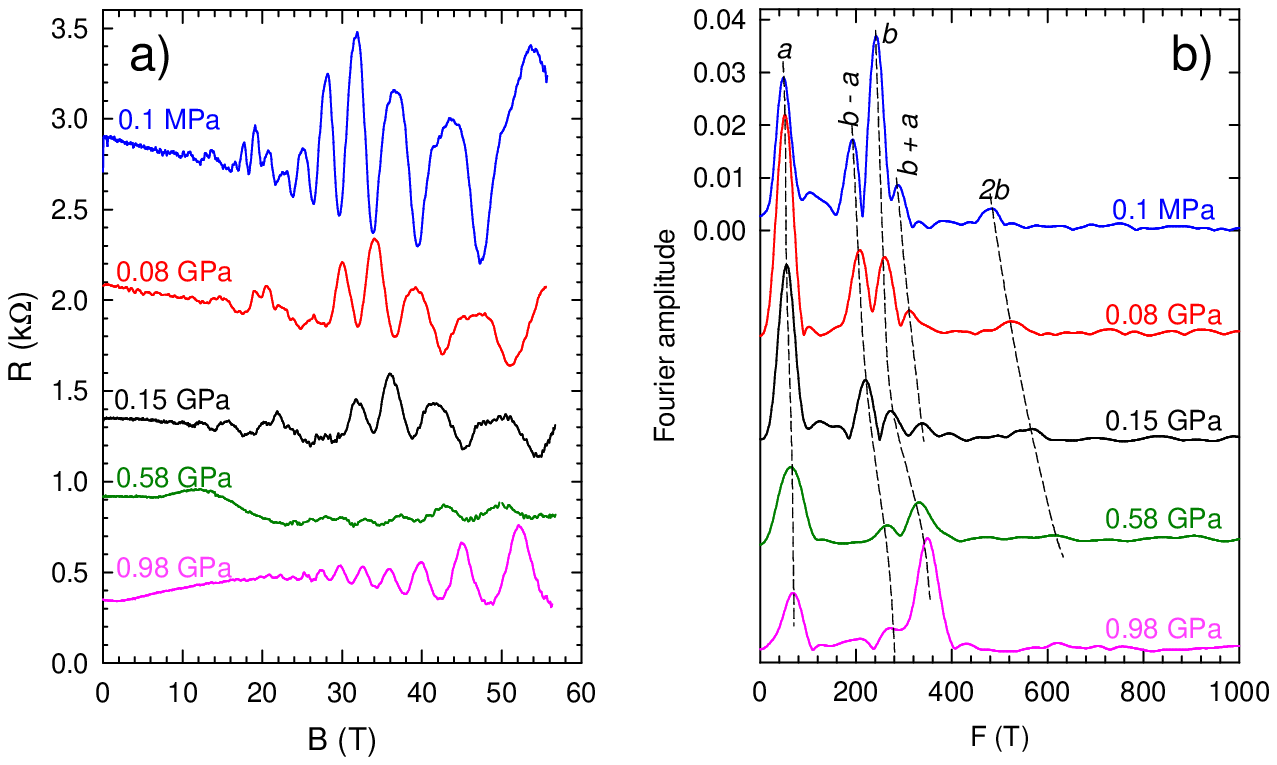}} \caption{\label{fig_rbtf} (Color
online) (a) Magnetoresistance at 1.6 K for the various pressures
studied. Data at 0.15 GPa and 0.98 GPa have been shifted down by
0.4 k$\Omega$ for clarity. (b) Fourier analysis of the oscillatory
magnetoresistance data deduced from Fig. \ref{fig_rbtf}(a).  }
\end{figure}

In agreement with data of Ref. \onlinecite{Au04}, the interlayer
zero-field resistance exhibits a pronounced minimum at T$_{min}$ =
27 K at ambient pressure (see Fig. \ref{fig_rt}). This behavior is
at variance with that reported for the in-plane resistance which
is metallic down to low temperature \cite{Pr03}. Even though the
temperature dependence of the interlayer resistance of
NH$_4$-Cr$\cdot$DMF remains qualitatively the same in the studied
pressure range (up to 1 GPa) \cite{Vi06}, the resistance minimum
of NH$_4$-Fe$\cdot$DMF is linearly shifted towards low
temperatures as the applied pressure increases (see the upper
inset of Fig \ref{fig_rt}). At 0.98 GPa, T$_{min}$ is decreased
down to 3 K and a strongly negative curvature is even observed
around 40 K at this applied pressure. In addition, the pressure
dependence of the interlayer resistance measured at room
temperature is d(lnR)/dP = (-1.35 $\pm$ 0.15) GPa$^{-1}$. This is
close to the value reported for NH$_4$-Cr$\cdot$DMF [d(lnR)/dP =
-1 GPa$^{-1}$] \cite{Vi06}. Otherwise, the resistance drop
observed below 1.8 K at ambient pressure, possibly connected with
the onset of a superconducting transition \cite{Au04}, is
suppressed from 0.08 GPa, as evidenced in the lower inset of Fig.
\ref{fig_rt}.

Magnetoresistance data collected at 1.6 K are displayed in Fig.
\ref{fig_rbtf}(a). It can be remarked that the background
magnetoresistance, which is slightly negative at low pressure, is
positive at 0.98 GPa while it exhibits a non-monotonic behavior at
0.58 GPa. Namely, a bump is observed around 12 T at this applied
pressure, in the studied temperature range from 1.6 K to 4.2 K.
The Fourier analysis of the oscillatory part of the
magnetoresistance data is displayed in Fig. \ref{fig_rbtf}(b). In
agreement with data of Ref. \onlinecite{Au04}, five frequencies
are observed at ambient pressure, namely, F$_a$ = (49 $\pm$ 2) T,
F$_{b - a}$ = (193 $\pm$ 2) T, F$_b$ = (241 $\pm$ 5) T, F$_{b +
a}$ = (287 $\pm$ 5) T and F$_{2b}$ = (482 $\pm$ 20) T. At first
sight, the only noticeable feature regarding Fourier analysis of
Fig. \ref{fig_rbtf}(b) is the pressure-induced vanishing of the
amplitude of the components at F$_{b + a}$ and F$_{2b}$. Recall
that, owing to the temperature and field dependencies of their
small amplitude, these two latter components were attributed
\cite{Au04} to frequency combinations typical of networks of
coupled orbits \cite{ClBr,Pi62,cf} rather than SdH oscillations
linked to either individual or magnetic breakdown (MB) orbits. As
previously reported, the relationship F$_a$ + F$_{b-a}$ = F$_b$,
which accounts for the compensation of these orbits, is observed
since F$_a$ + F$_{b-a}$ = (242 $\pm$ 4) T.

\begin{figure}                                                      
\centering \resizebox{0.8\columnwidth}{!}
{\includegraphics*{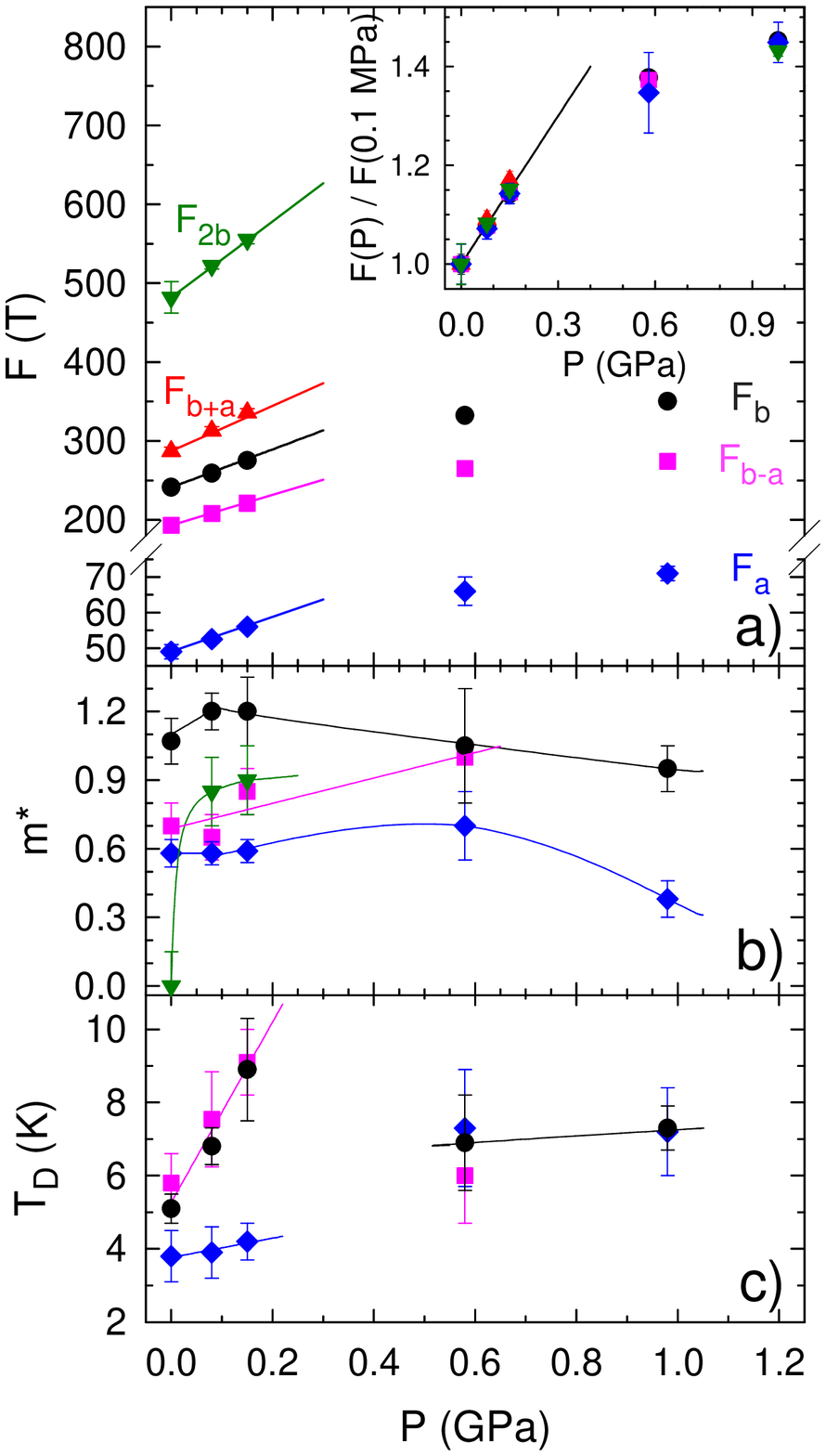}}
\caption{\label{fig_F_mc_TD(P)} (Color online) Pressure dependence
of (a) the various Fourier components observed in the oscillatory
magnetoresistance, (b) the effective masses and (c) the Dingle
temperatures (see text).  The inset of (a) displays the pressure
dependence of the frequencies normalized to their ambient pressure
value. Solid straight lines in (a) correspond to F(P) / F(0.1 MPa)
= 1 + $\kappa$P with $\kappa$ = 1 GPa$^{-1}$. Solid lines in (b)
and (c) are guides to the eye. }
\end{figure}

As the applied pressure increases, the Fourier spectra remain
similar. In particular, the above mentioned linear relation is
still valid at high pressure since, e. g. at 0.98 GPa, F$_a$ = (71
$\pm$ 2) T, F$_{b - a}$ = (274 $\pm$ 6) T [which yields F$_a$ +
F$_{b-a}$ = (345 $\pm$ 8) T] and F$_b$ = (350 $\pm$ 1) T. To be
more exact, the pressure dependence of the relative value of the
frequencies F(P)/F(P = 0.1 MPa) is the same for all the Fourier
components observed, as evidenced in the inset of Fig.
\ref{fig_F_mc_TD(P)}(a). This result is not only at variance with
the behavior of NH$_4$-Cr$\cdot$DMF, of which the FS topology is
strongly modified under pressure \cite{Vi06}, but also with the
data for compounds that illustrate the linear chain of coupled
orbits model\cite{Ca94,Ka95,Br95}. Indeed, the area of the closed
$\alpha$ orbit of all these compounds is significantly more
sensitive to the applied pressure than that of the MB-induced
$\beta$ orbit, although the FS topology remains qualitatively the
same as the applied pressure varies.  In this respect, it can be
noticed that the area of the $\beta$ orbit is equal to that of the
FBZ, just as it is the case of the $\bigodot$ orbit (see Fig.
\ref{fig_sf}). Unfortunately, it was not possible to observe SdH
oscillations linked to this latter orbit in the present case.
Nevertheless, it is unlikely that the pressure dependence of the
observed FS pieces area follows that of the FBZ. Indeed the
observed huge increase of frequency, namely about 45 percent at
0.98 GPa, as reported in the inset of Fig.
\ref{fig_F_mc_TD(P)}(a), cannot reflect that of the FBZ area in
view of the compressibility values reported for organic compounds.
The influence of the applied pressure on the FS topology of some
organic metals were successfully modelled by a modification of
selected molecular orbital interactions \cite{Ca94, Au95, Ca96}.
In order to reproduce the pressure effect observed in the present
case, simulations of the FS were carried out, in which selected
transfer integrals were varied. However, any attempt of such tight
binding band structure calculations, which in most cases induces a
slight rotation of the $\bigodot$ orbit \cite{Ro04}, fails to
reproduce the experimental data. More precisely, the cross section
of the electron and of one of the two hole tubes can significantly
increase but, in any case, the cross section of the other hole
tube decreases. In none of the simulations a significant and
simultaneous increase of the three cross sections was observed
even though electron-hole orbits compensation always holds.\\

Let us consider now the temperature and field dependence of the
amplitude of the oscillations as the pressure varies. This can be
achieved in the framework of the Lifshits-Kosevich (LK)
formula\cite{Sh84} which has been reported to satisfactorily
account for the data of SdH oscillations linked to closed orbits
in q-2D networks of compensated electron and hole orbits, even in
the case of crystals with low scattering rate \cite{ClBr}.
Effective masses (m$^*_i$), deduced from the temperature
dependence of the amplitudes are given in Fig.
\ref{fig_F_mc_TD(P)}(b). Since magnetoresistance data at ambient
pressure, which are in agreement with those of Ref.
\onlinecite{Au04}, were only recorded at 1.6 K and 4.2 K, the
effective masses given in Fig. \ref{fig_F_mc_TD(P)}(b) are taken
from Ref. \onlinecite{Au04}. Despite the large error bars obtained
for the data at 0.58 GPa (at this pressure, the amplitude of the
oscillations is rather small, as displayed in Fig. \ref{fig_rbtf})
it can be concluded that large variations of the effective masses
occur as the applied pressure varies. It can also be noticed that
m$^*_{2b}$ has finite values under pressure, even though they are
significantly lower than expected within the semiclassical picture
(which predicts m$^*_{2b}$ = 2m$^*_{b}$) as already observed in
many 2D organic conductors \cite{Sa96Uj97Ha98}. This is at
variance with data at ambient pressure for which a
temperature-independent amplitude has been reported, which is
therefore compatible with a zero-effective mass \cite{Au04}. This
feature, which can be considered in connection with the above
mentioned pressure-induced decrease of the amplitude of the
Fourier component with the frequency F$_{b+a}$, indicates that the
non-semiclassical features of the oscillatory spectra vanish as
the applied pressure increases. Since nothing is known regarding
the value of the MB gaps between the orbits, MB was not considered
in the analysis of the field-dependent amplitude of the
oscillations. The deduced Dingle temperatures are given in Fig.
\ref{fig_F_mc_TD(P)}(c). As already reported for the salts of this
family with the unsymmetrical DMF solvent \cite{Au04,Vi06}, large
Dingle temperatures are observed. In the low pressure range, up to
0.15 GPa, different values of the Dingle temperature are obtained
for $a$, on the one hand, and $b-a$ and $b$ orbits, on the other
hand. A jump of T$_D$ is observed between 0.15 GPa and 0.58 GPa
and, contrary to the data at low pressure, the same Dingle
temperatures are observed, within the error bars, for $a$ and $b$
at 0.58 GPa and 0.98 GPa. Even though the pressure dependence of
the observed frequencies indicates that the FS topology remains
qualitatively unchanged as the pressure varies, the large pressure
dependence of the effective masses, the non-monotonous behavior of
the Dingle temperatures, as well as the behavior of the
magnetoresistance under pressure [see Fig. \ref{fig_rbtf}(a)]
suggest variations of the electronic structure
as the applied pressure varies.\\

In summary, the most striking feature regarding the SdH
oscillatory spectrum of NH$_4$-Fe$\cdot$DMF under pressure is the
very strong pressure sensitivity of the observed frequencies which
increase by 45 percent at 0.98 GPa. In line with this result, a
sizeable pressure dependence of the zero-field interlayer
resistance is observed. Nevertheless, the pressure dependence of
the normalized value of the various frequencies is the same,
within the error bars, for all of them which suggests that the FS
topology remains qualitatively the same in the applied pressure
range studied. Even more, it is consistent with the preservation
of the orbits compensation as the applied pressure varies. This
behavior is at variance with that of both the related compound
NH$_4$-Cr$\cdot$DMF \cite{Vi06} and the organic compounds that
illustrate the model of linear chains of orbits
\cite{Ca94,Ka95,Br95}. Oppositely, the behavior of the zero-field
resistance as the temperature varies is significantly modified by
applied pressure. In addition, significant variations of the
effective masses and non-monotonous behaviors of the scattering
rates linked to the various Fourier components are observed under
pressure. These features suggest that, despite the above discussed
pressure dependence of the frequencies, some variations of the
electronic structure occur under pressure. The present results
clearly demonstrate that we are still far from completely
understand the subtle details of the electronic structure of this
remarkable family of organic compounds.\\

\begin{acknowledgments}
 This work was supported by the French-Spanish exchange
 programm between CNRS and CSIC (number 16 210), Euromagnet under the European Union contract
 R113-CT-2004-506239, DGI-Spain (Project BFM2003-03372-C03) and by Generalitat de Catalunya (Project 2005 SGR
 683). Helpful discussions with G. Rikken are acknowledged.
\end{acknowledgments}

$\dagger$ author for correspondence: audouard@lncmp.org

 \bibliography{apssamp}

 \end{document}